# A comparison study of CNN denoisers on PRNU extraction

Hui Zeng [a, b#], Morteza Darvish Morshedi Hosseini [b#], Kang Deng [a], Anjie Peng [a], Miroslav Goljan [b*]

[a] *School of Computer Science & Technology, Southwest University of Science and Technology, Mianyang, China*
[b] *Department of ECE, SUNY Binghamton, NY, USA*

**Abstract** Performance of the sensor-based camera identification (SCI) method heavily relies on the denoising filter in estimating Photo-Response Non-Uniformity (PRNU). Given various attempts on enhancing the quality of the extracted PRNU, it still suffers from unsatisfactory performance in low-resolution images and high computational demand. Leveraging the similarity of PRNU estimation and image denoising, we take advantage of the latest achievements of Convolutional Neural Network (CNN)-based denoisers for PRNU extraction. In this paper, a comparative evaluation of such CNN denoisers on SCI performance is carried out on the public "Dresden Image Database". Our findings are two-fold. From one aspect, both the PRNU extraction and image denoising separate noise from the image content. Hence, SCI can benefit from the recent CNN denoisers if carefully trained. From another aspect, the goals and the scenarios of PRNU extraction and image denoising are different since one optimizes the quality of noise and the other optimizes the image quality. A carefully tailored training is needed when CNN denoisers are used for PRNU estimation. Alternative strategies of training data preparation and loss function design are analyzed theoretically and evaluated experimentally. We point out that feeding the CNNs with image-PRNU pairs and training them with correlation-based loss function result in the best PRNU estimation performance. To facilitate further studies of SCI, we also propose a minimum-loss camera fingerprint quantization scheme using which we save the fingerprints as image files in PNG format. Furthermore, we make the quantized fingerprints of the cameras from the "Dresden Image Database" publicly available.

**Keywords** Camera identification · Photo-Response Non-Uniformity · denoising filter · convolutional neural network · fingerprint quantization

## 1 Introduction

Photo-Response Non-Uniformity (PRNU)-based method has dominated the camera identification research since it was proposed more than a decade ago [1]. Unlike other feature-based methods [2, 3], the method based on PRNU can identify not only camera models but also individual devices of the same model [4, 5]. PRNU, mainly caused by the inhomogeneity of silicon wafers [6], can be estimated from a single image or even a video frame using a denoising filter and represented in the form of a noise matrix. Averaging a number of PRNU estimates from images produced by the same camera can generate a camera fingerprint. A similarity between the PRNU from one probe image and a camera fingerprint is an essential clue for deciding whether the camera took this probe image or not.

Concerning the reliability of the sensor-based camera identification, the denoising filter used for extracting the PRNU plays a key role. In [1, 7], a Discrete Wavelet Transform (DWT)-based denoising filter [8] was used to obtain an estimate of PRNU. Ever since then, several advanced denoising filters have been adopted to extract PRNU. In [9], PRNU was extracted with a curvelet transform-based denoising filter, which declared better performance for images including highly textured regions. In [10], an edge-adaptive PRNU predictor based on context-adaptive interpolation (CAI) and pixel-wise adaptive Wiener filter was proposed. Context-adaptive guided image filtering was utilized for PRNU extraction in [11]. In [12], the BM3D filter [13] that exploits the non-local self-similarity of images was used to improve PRNU extraction. While its performance is excellent, the BM3D based method suffers from a much higher computational demand than the other methods mentioned above. In [14], Dual-Tree Complex Wavelet Transform (DTCWT) was used as an alternative to DWT in PRNU extraction. Thanks to its rich directional selectivity and shift-invariance property, DTCWT overperforms DWT with a comparable computational time.

Considering the numerous attempts to boost the PRNU extraction performance, data-driven methods are still lacking. In [15], the authors fed corresponding image-camera fingerprint pairs to Convolutional Neural Network (CNN) and reported promising results when the neural networks were trained separately for each camera. However, training individual CNN models for every probe camera is prohibitively expensive, even impossible in practice, because a large number of training images from every camera may be unavailable. In this paper, we make an extensive study of the feasibility of training a universal CNN model for PRNU extraction. Specifically, we select four state-of-the-art CNN denoisers and train them for the universal PRNU extraction purpose. Then, the PRNU estimation performance is compared in terms of correlation metrics and ROC curves. We also examine the effect of training strategies on the PRNU extraction performance. Experiments show that training a universal model based on CNN denoisers is feasible if we consider PRNU specifics in training.

We point out fundamental differences between the camera identification problem and the similar *camera model* identification problem: having different objectives, solutions rely on different physical artifacts. Camera model identification leverages the so-called Non-Unique Artifacts (NUA) in Sensor Pattern Noise (SPN) to determine the source camera model of a probe image. In contrast, camera identification focuses on extracting the unique PRNU signal that can distinguish individual cameras, even of the same make and model. Several deep learning-based works focus on camera model identification only [16, 17].

The contributions of this work are as follows:

1. We present a comparative study of the CNN denoisers for the PRNU estimation purpose. We show that the SCI can indeed benefit from the recent advances of CNN denoisers, yet a more advanced CNN denoiser does not necessarily mean a better PRNU extractor.

2. By exploring characteristics of PRNU, we show that tailored training of the CNN denoisers can improve the PRNU estimation performance. Specifically, we propose a correlation-based loss function that outperforms the MSE-based loss function in adapting CNN denoisers for PRNU estimation.

3. Last but not least importantly, we propose a camera fingerprint quantization scheme that can significantly reduce the required storage requirements for camera fingerprints. Using the proposed scheme, we make the quantized fingerprints of the cameras from the Dresden Image Database publicly available, which can significantly relieve researchers from the repeated and tedious work needed for camera fingerprints generation.

The rest of this paper is organized as follows. In the next section, we briefly review the process of sensor-based camera identification and introduce an efficient fingerprint quantization scheme. The third section introduces four candidate CNN denoisers and alternative training strategies for PRNU estimation. Experimental results are presented and analyzed in Section 4. Section 5 concludes the paper.

**2 Sensor-based camera identification and fingerprint quantization**

2.1 Sensor-based camera identification

The PRNU matrix, denoted as $W$, is estimated from a given image $I$ as the difference between the image and its denoised version,

$$W = I - F(I), \qquad (1)$$

where $F$ refers to the denoising filter. Signal processing techniques are then used to suppress unwanted artifacts from $W$, such as row and column average biases, periodic signals, and peaks in the frequency domain [7, 18]. The fingerprint $K$ of a given camera can then be estimated from several images from this camera by averaging over $W^{(i)}$ from the set of images $I^{(i)}, i = 1, 2, \dots, N$ [1] or using a Maximum-Likelihood Estimator (MLE) [7]:

$$K = \frac{\sum_{i=1}^{N} W^{(i)} I^{(i)}}{\sum_{i=1}^{N} \left(I^{(i)}\right)^2}, \qquad (2)$$

where $N$ is the number of images used to estimate $K$. A correlation statistic, such as the normalized cross-correlation or peak-to-correlation-energy (PCE) [19], is used to measure the similarity between $W_p$ (the PRNU of a probe image $I_p$) and $K_A$ (the fingerprint of the reference camera $A$). The normalized cross-correlation between two variables $x$ and $y$ is defined as

$$\rho(x, y) = \frac{(x - \bar{x}) \cdot (y - \bar{y})}{||x - \bar{x}|| \, ||y - \bar{y}||}, \qquad (3)$$

where $\bar{x}$ and $\bar{y}$ denote the mean value of $x$ and $y$, respectively, and $\cdot$ denotes the dot product. The PCE is defined as

$$PCE = \frac{(|s|-|\Omega|)\rho_{(0,0)}^2}{\sum_{s \notin \Omega} \rho_s^2} \cdot sign(\rho_{(0,0)}) \qquad (4)$$

with $s$ being the two-dimensional index of $\rho$. $\Omega$ represents a small region around $\rho(0, 0)$ and $|\Omega|$ is the number of elements in $\Omega$. The probe image $I_p$ is declared to be from camera $A$ when the PCE value is greater than a given threshold $\tau$.

2.2 Fingerprint quantization

In the traditional SCI scheme, once the camera fingerprint is estimated with (2), it is stored with double or single precision. However, when the number of fingerprints to store or share is huge, the storage requirement or communication channel bandwidth poses practical limitations. As a consequence, researchers tend to regenerate the camera fingerprints once they are needed, which becomes a time-consuming and tedious process itself.

In [20], the authors proposed to binarize the camera fingerprints to overcome this problem. However, binarization comes with a significant loss for the correlation-based detector performance. Here, we propose a uniform quantization scheme with minimum loss in SCI performance. Specifically, a camera fingerprint is quantized to 8 bits per sample that can be stored as a grayscale PNG image, allowing an easy cross-platform exchange of fingerprint data. The proposed quantizer can be expressed as

$$Q(\mathbf{K}, a) = \max(\min(round(a\mathbf{K} - 0.5) + 128, 255), 0), \qquad (5)$$

where $a$ is the scaling parameter. We experimentally determine $a = 32.5$ for quantizing camera fingerprints to maximize correlation $\rho\left(\mathbf{K}, Q(\mathbf{K}, a)\right)$. With this quantization scheme, we can save 75% (compared to single-precision) or 87.5% (compared to double-precision) storage consumption without losing correlation-detector performance, as will be verified in Section 4.2.

**3 PRNU extraction with CNN denoisers**

Although the existing sensor-based camera identification methods have made significant progress in the last decade, their performance is bounded by the denoising filter in (1). Consequently, the camera identification results for strongly textured images or small image blocks are less reliable. From another perspective, the last decade witnessed the rapid progress of deep learning-based image denoising methods [21-26]. In [21], a multi-layer perceptron (MLP) was successfully applied for image denoising. It was for the first time that a data-driven image denoising method achieved performance comparable with BM3D. In [22], Zhang et al. borrowed the structure of VGGnet [27] to design an image denoising network and called it DnCNN. DnCNN achieved superior denoising performance to BM3D and has drawn much attention since it was proposed. In that scheme, however, a denoising model needs to be learned for each noise level. In [23], the same authors of [22] proposed a flexible denoising CNN, namely FFDNet, with a tunable noise level map as the input. FFDNet can be faster than DnCNN as it operates on downsampled images. In [24], an attention mechanism was applied to a CNN to improve denoising performance further.

While the above CNN denoisers aim to restore images corrupted with additive white Gaussian noise (AWGN), some more recent works try to handle real noisy images, especially high ISO images. In [25], a convolutional blind denoising network was proposed for denoising real noisy images by first estimating the noise level with one subnet and then denoising with another subnet. In [26], to overcome the lack of training pairs in denoising real noisy images, a generative adversarial network (GAN) was used for generating clean–noisy image pairs.

A survey of deep learning methods for image denoising was presented in [28]. The close relationship between image denoising and PRNU estimation, as shown in (1), motivated us to adapt CNN denoisers for PRNU extraction. After an overall evaluation of the image denoising performance and efficiency, we select four representative CNN denoisers and evaluate their performance for PRNU extraction.

3.1 Selected CNN denoisers

3.1.1 DnCNN

DnCNN is one of the most popular CNN denoisers. As shown in Fig. 1(a), there are three types of layers for a DnCNN with depth $D$. The first layer includes a convolutional component (Conv) followed by rectified linear units (ReLU) [29]. For layers $2 \sim (D - 1)$, batch normalization [30] is added between Conv and ReLU. For the last layer, Conv is used to reconstruct the output. Zero-padding is employed in convolutions to keep the size of feature maps unchanged throughout the network. As shown in the rightmost part of Fig. 1(a), the DnCNN uses the residual image rather than the noise-free image as the training target. The MSE based loss function $L$ can be formalized as

$$L(\boldsymbol{\theta}) = \frac{1}{2N}\sum_{i=1}^{N}\|F(\boldsymbol{I}_i;\boldsymbol{\theta}) - (\boldsymbol{I}_i - \boldsymbol{x}_i)\|^2, \qquad (6)$$

where $\boldsymbol{\theta}$ represents the set of network parameters. It was reported that the integration of residual learning [31] and batch normalization could speed up the training process and boost denoising performance.

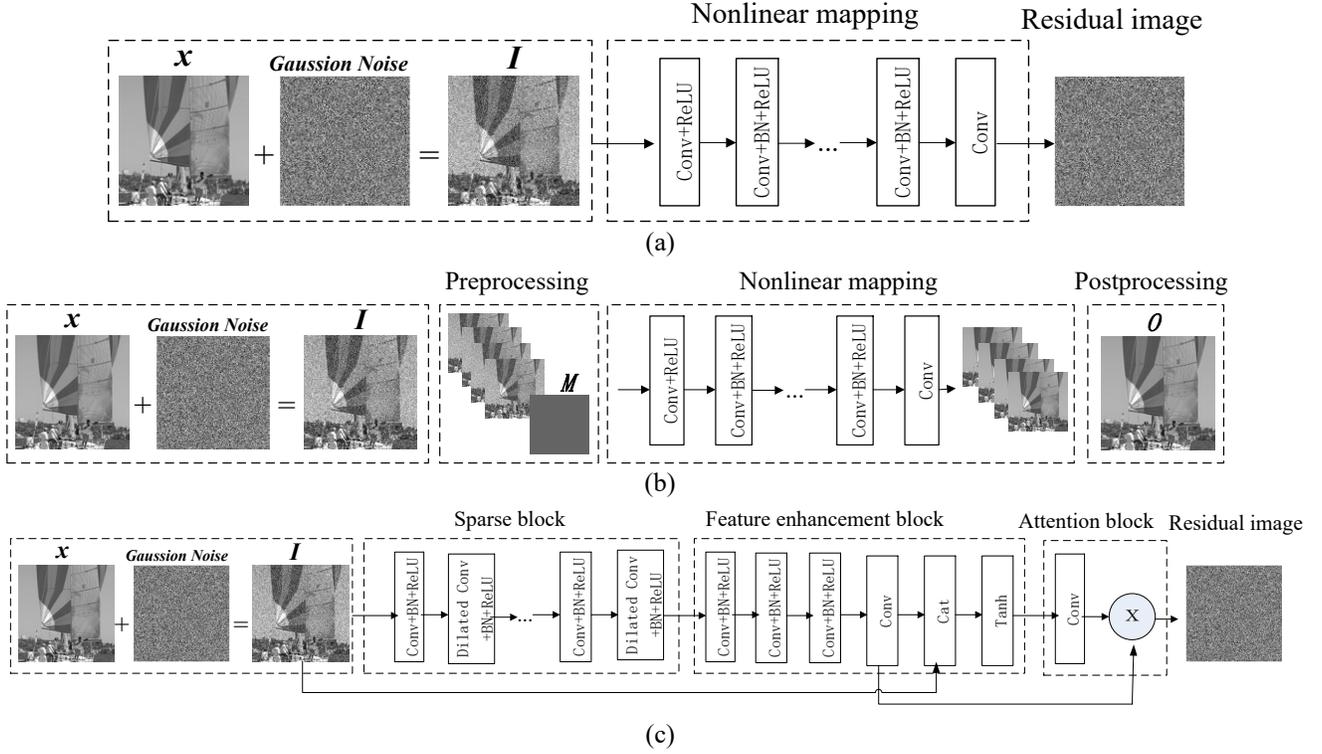

**Fig. 1.** The architecture of the CNN denoisers evaluated in this paper. (a) DnCNN, (b) FFDNet, (c) ADNet.

3.1.2 FFDNet

The second CNN denoiser to evaluate is FFDNet, with its architecture illustrated in Fig. 1(b). It includes three stages according to their functions, namely preprocessing, nonlinear mapping, and postprocessing. Let the input image $\boldsymbol{I}$ be a grayscale image with the size of $H \times W$ pixels. In the preprocessing stage, $\boldsymbol{I}$ is down-sampled into four sub-images with the size of $H/2 \times W/2$ pixels. After that, a noise level map $\boldsymbol{M}$ with the same size of the sub-images is concatenated to the four sub-images. For spatially invariant AWGN with noise level $\sigma$, $\boldsymbol{M}$ is a uniform map with all elements being $\sigma$, which is the case of our study. To summarize, $\boldsymbol{I}$ is reorganized as a tensor $\boldsymbol{y}$ with the size of $H/2 \times W/2 \times 5$. The nonlinear mapping stage of FFDNet is the same as that of DnCNN. In the postprocessing stage, the output of the last convolution layer, having the size of $H/2 \times W/2 \times 4$, is upsampled into a denoised image $\boldsymbol{O}$ with the size of $H \times W$ pixels. The most significant difference between FFDNet and DnCNN is that FFDNet takes the noise level map $\boldsymbol{M}$ as the input, enabling a trained FFDNet model to handle noisy images for a wide range of noise levels and be much more flexible than DnCNN. Unlike a fixed $\sigma$ that is used in training a specific DnCNN model, a range of $\sigma$ from the interval $[0, \sigma_{max}]$ is used in training an instance of FFDNet.

3.1.3 ADNet

ADNet is composed of a sparse block, a feature enhancement block, and an attention block, as shown in Fig. 1(c). The sparse block is a stack of standard and dilated convolutional layers that can enlarge the receptive field size without making the network too deep. The feature enhancement block can integrate the global and local features. The attention block can capture the key noisy features hidden in the complex background. ADNet, with its elaborated design, can outperform DnCNN in image denoising in terms of peak signal-to-noise ratio (PSNR) [24].

3.1.4 DANet

The last CNN denoiser to evaluate is the DANet. It is composed of three parts: denoiser, generator, and discriminator. The generator and the discriminator are designed for generating clean–synthetic noisy image pairs to overcome the lack of training pairs in denoising real noisy images. The denoiser itself utilizes the U-Net [32] backbone. A detailed description of its architecture is out of the scope of this paper. Hence, we refer interested readers to the original publication [26] for more details. DANet reportedly achieves excellent denoising performance on real noisy images, e.g., high ISO images. However, the adversarial training process makes the training stage of DANet very time-consuming.

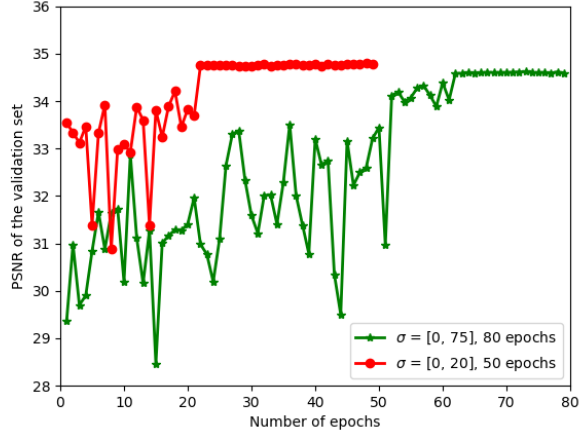

**Fig. 2.** Denoising performance comparison on the Set12 validation set when the FFDNet is trained with different settings. The green line denotes the default setting and the red line denotes our used setting. The used setting converges faster and performs better when the standard deviation of the Gaussian noise added to the validation set equals 10.

3.2 Training strategies

In this subsection, we discuss the training strategies from two aspects: training data preparation and loss function design, which significantly affect PRNU estimation performance.

When it comes to preparing input-target image pairs for training a CNN denoiser, two conventional approaches exist. The first one adds synthetic noise to noise-free images [22, 23, 24]. The second one uses real noisy images and their noise-free counterpart [25, 26]. The first strategy is cheap to implement, yet it is restricted to the prior model of the synthetic noise, e.g., AWGN ( $N(0, \sigma^2)$ ). The second approach is closer to the real noise in images, but the process of estimating the corresponding noise-free counterparts is usually very expensive. We intentionally choose representative CNN denoisers of both strategies in the comparison study in Section 4.

When such CNN denoisers are adapted for PRNU extraction, it is necessary to bear in mind that the PRNU signal is very weak compared to the image signal [1]. Hence, $\sigma$ should be carefully set to a small value in training the CNNs. For DnCNN and ADNet, we follow [15] and set $\sigma=3$ in this study. In training FFDNet, the range of $\sigma$ is set to [0, 20], rather than the default setting of [0, 75]. To justify this choice, we present a comparison of the training process of different noise level settings. As shown in Fig. 2, focusing on the weak noise scenario helps the CNN converge faster in training and perform slightly better when denoising images that are corrupted by weak noise. The training dataset is the Waterloo Exploration Database [33], consisting of 4,744 natural images in PNG format, most of which are around 0.2 M pixels. The validation set is the Set12 dataset, a collection of widely-used test images, such as "Cameraman", "Lena", and "Barbara".

We additionally consider a third strategy in preparing input-target image pairs for training a PRNU extractor. Similar to [15], an image from a certain camera is used as the noisy input, and the reference fingerprint of this camera serves as the target. In Section 4, we experimentally verify that the third strategy yields an improved PRNU estimation performance.

In the context of image denoising, all of the mentioned CNN denoisers above are trained with the MSE loss function as (6), i.e., forcing the network output to be *close* to the target. However, the goal of SCI is different. The extracted PRNU from the probe image is preferred to be *similar* to the reference PRNU. The following toy example makes the difference more clear. Suppose there is a four-pixel camera whose reference PRNU is

$$K = \begin{bmatrix} 0.2 & 0.2 \\ -0.2 & -0.2 \end{bmatrix}$$

and there are two estimates of PRNU, $\widehat{K}_1$ and $\widehat{K}_2$, output from a CNN denoiser.

$$\widehat{K}_1 = \begin{bmatrix} 1 & 1 \\ -1 & -1 \end{bmatrix}, \quad \widehat{K}_2 = \begin{bmatrix} -0.1 & 0.1 \\ -0.1 & 0.1 \end{bmatrix}$$

Under the MSE criteria, $\widehat{K}_2$ is a better estimation of $K$ than $\widehat{K}_1$ because $\|\widehat{K}_2 - K\| < \|\widehat{K}_1 - K\|$. However, if we examine the correlation-based similarity, $\widehat{K}_1$ becomes a better estimation than $\widehat{K}_2$, since $\rho(\widehat{K}_1, K) = 1 > \rho(\widehat{K}_2, K)$. Hence, for PRNU extraction, it is necessary to utilize the $\rho$-based loss function instead of the MSE loss function. The loss based on $\rho$ encourages the network to estimate a PRNU as similar as possible to the provided reference PRNU. Knowing that $|\rho| \leq 1$ and considering the convention of positive loss, we define the $\rho$-based loss function as

$$L = 1 - \rho(\mathbf{z}, \mathbf{K}), \tag{7}$$

where $\mathbf{z} = F(\mathbf{I}; \boldsymbol{\theta})$ is the output of the network with the set of parameters $\boldsymbol{\theta}$. The backward function of the $\rho$-based loss function can be derived from (3) as

$$\frac{\partial L}{\partial \mathbf{z}} = \frac{((\mathbf{z}-\bar{\mathbf{z}}) \cdot (\mathbf{K}-\bar{\mathbf{K}}))\mathbf{z} - \|\mathbf{z}-\bar{\mathbf{z}}\|^2 \mathbf{K}}{\|\mathbf{z}-\bar{\mathbf{z}}\|^3}, \tag{8}$$

In Section 4, the performance of PRNU estimation with MSE loss and $\rho$-based loss is compared and analyzed.

## 4 Experimental Results

We start this section with the experimental setup and verification of the effectiveness of the camera fingerprint quantization scheme. After that, four state-of-the-art CNN denoisers, i.e., DnCNN, FFDNet, ADNet, and DANet, are compared in terms of PRNU estimation performance. Taking the DnCNN as an example, we then compare different training strategies for PRNU estimation.

4.1 Experimental setup

For the training dataset of the CNN denoisers, we try our best to follow the settings of the original publications. For training DnCNN, 400 images of size 180×180 selected from ImageNet [34] are used. For training FFDNet and ADNet, the Waterloo Exploration Database [33] is used. The standard deviation of the synthetic noise for training DnCNN and ADNet is set to 3 according to the recommendation of [15]. The range of the standard deviation of the synthetic noise for training FFDNet is set as [0, 20] as discussed in Section 3. The dataset used for training FFDNet and ADNet is much larger than that for training DnCNN, possibly causing different PRNU extraction performances, as discussed later in this section. For the DANet, the training time is prohibitive for us. Thus we use the trained model provided by its authors.

The testing dataset uses 40 cameras of 11 models from the publicly available "Dresden Image Database" [35]. Table 1 shows the camera model, camera individual, and the number of probe images. All the images are in JPEG format. To reduce the effect of sub-optimal filtering in generating the reference camera fingerprints, for the cameras that flat-field images are available, we calculate the camera fingerprint from 25-50 flat-field images using the DWT method. For the cameras that flat-field images are not available, the camera fingerprints are calculated from 50 natural images. Other 150 images with natural scenes are taken as probe images, except for Pentax_OptioA40 C3, with only 114 images available. Most of the existing SCI methods achieve nearly perfect performance on large resolution images. In order to be able to see the performance differences of various SCI methods, image blocks with the size of 128×128 pixels and 64×64 pixels are used in this study. For such small image samples, detection errors occur much more often. To prepare the data samples, four images with the size of 128×128 (or 64×64) are cropped from the center of each original image. Once a camera is fixed as the reference camera, the images captured with this camera (at the corresponding location) are used as positive samples, and the images from three other cameras are used as negative samples. The SCI results of all cameras (and all locations) are merged in this study. This way, we obtain $(39 \times 150 + 114) \times 4 = 23856$ positive samples and $23856 \times 3 = 98568$ negative samples.

**Table 1** Images used in the experiments.

| Camera model | Camera individual | Number of probe images | Camera model | Camera individual | Number of probe images |
| --- | --- | --- | --- | --- | --- |
| Olympus_1050SW | C0-C4 | 750 | Rollei_RCP | C0-C2 | 300 |
| Panasonic_FZ50 | C0-C2 | 450 | Kodak_M1063 | C0-C4 | 750 |
| Nikon_D200 | C0-C1 | 300 | Samsung_L74 | C0-C2 | 450 |
| Pentax_OptioA40 | C0-C3 | 554 | Samsung_NV15 | C0-C2 | 450 |
| Praktica_DCZ5.9 | C0-C4 | 750 | Sony_DSC-H50 | C0-C1 | 300 |
| Ricoh_GX100 | C0-C4 | 750 | Total | | 5964 |

**Table 2** Average PCE values of the seven methods for positive samples of size 128×128 with different PRNU quantization schemes. The best results are highlighted in **bold**.

|  | Data-driven | | | | Non–data-driven | | |
| --- | --- | --- | --- | --- | --- | --- | --- |
|  | DnCNN | FFDNet | DANet | ADNet | DWT | DTCWT | BM3D |
| Single precision | 13.6 | **16.5** | 9.6 | 15.1 | 13.5 | 15.3 | 15.8 |
| Quantized fingerprint | 13.6 | **16.4** | 9.6 | 15.0 | 13.3 | 15.5 | 15.6 |
| Binarized fingerprint [20] | 9.1 | **10.9** | 7.2 | 10.0 | 8.9 | 10.4 | 10.5 |

4.2 Empirical evaluation of quantized camera fingerprints

We proceed with PRNU signal extraction with four CNN denoisers for each probe image. Then, for each extracted PRNU, we calculate its PCE values with the single-precision camera fingerprint, the proposed quantized camera fingerprint, and the binarized camera fingerprint. As can be observed from Table 2, the proposed quantization scheme hardly affects the correlation detector performance while saving up to 75% (compared to single precision) storage consumption. By comparison, although the fingerprint binarization scheme can further reduce storage consumption, it comes at the expense of a drop in PRNU matching performance. Take the DWT method for example, the average PCE value for positive samples reduces by 34%, from 13.5 to 8.9, after binarization. Hence, we use the quantized camera fingerprints for the rest of this paper.

4.3 Comparison of CNN denoisers

To compare the four CNN denoisers that are trained using their original training strategies, we again refer to the average PCE values in Table 2. For completeness, the results of three non-data-driven methods are also listed. The denoising parameter $\sigma$ is set to 2 for the DWT method and 1.8 for the DTCWT method as recommended in [1, 14]. For the BM3D method, we set $\sigma = 5$ as recommended in [36]. For all the compared methods, the post-processings, including zero-meaning and Wiener filtering [7], are applied to the reference fingerprints only, not to the probe images. Higher PCE values indicate a better PRNU extractor. Table 2 shows that all the CNN denoisers except DANet achieve comparable or better PRNU estimation performance with the state-of-the-art non-data-driven methods. FFDNet has the highest average PCE value among the CNN denoisers, and ADNet ranks second. We conjecture that the much richer database used for training contributes their advantages over DnCNN. According to our experiments, the performance of image denoising and that of PRNU extraction are not necessarily consistent with each other. Especially, the poor performance of DANet in PRNU extraction is rather unexpected. We hypothesize that this is because the diversity of image scenes used for training DANet is very limited.

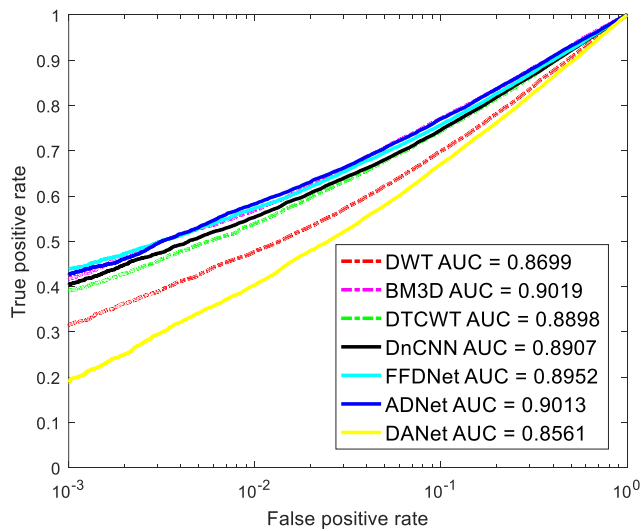

**Fig. 3.** Performance comparison of different PRNU estimation methods in terms of ROC curves. AUC is the Area under Curve.

**Table 3** Average PCE values from positive samples for different training strategies. The best results are highlighted in **bold**.

|  | DnCNN | DnCNN-SPN-MSE | DnCNN-SPN-Rho |
|---|---|---|---|
| 128 × 128 | 13.6 | **18.5** | **18.5** |
| 64 × 64 | 4.1 | 5 | **5.1** |

The comparison above does not take the distribution of PCE values of negative samples into account. ROC curves provide a more detailed evaluation of binary classification. We compare ROC curves on 128×128 images in Fig. 3. It reveals that ADNet achieves the best performance among the CNN denoisers in terms of AUC value that is also comparable with the best AUC achieved by non-data-driven methods. On the other hand, DANet shows the worst performance among all the compared methods, particularly in the low false alarm area. Take false positive rate FP = 0.001 for example, DnCNN, FFDNet, ADNet, and DANet achieve true positive rates TP = 0.404, 0.440, 0.426, and 0.188, respectively. Since five very different denoisers (two non-data-driven methods and three CNN ones) lead to similar PRNU extraction performance, it suggests that we are possibly approaching the upper bound of what is possible for the given data.

4.4 Comparison of training strategies

Considering the competitive performance and simple structure of DnCNN, we use it as the baseline to compare the PRNU estimation performance of different training strategies. We call the network trained with image-PRNU pairs and MSE loss function as DnCNN-SPN-MSE, and the network trained with image-PRNU pairs and $\rho$-based loss function as DnCNN-SPN-Rho. Note that DnCNN-SPN-MSE is similar to SPNCNN in [15], except that the authors in [15] train dedicated instances for each camera, whereas we train a universal model for all cameras. We admit that using a richer database in training may benefit the trained model. Thus, for a fair comparison, for all the alternative training strategies, we tightly follow the original DnCNN using the same patch size (40 by 40), a comparable number of patches in each epoch, and the same number of epochs (50) for training. To this end, only 150 original size images from 5 cameras (each camera responsible for 30 images) in VISION [37] are used for training DnCNN-SPN-MSE and DnCNN-SPN-Rho.

Table 3 reports the average PCE values for positive samples of the compared training strategies on both 128×128 images and 64×64 images. DnCNN-SPN-Rho ranks first, and a notable boost can be observed when DnCNN is trained with image-

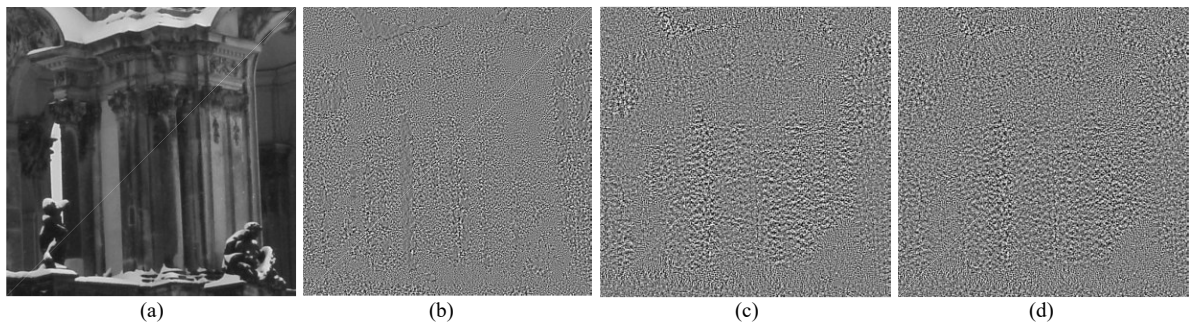

**Fig. 4.** Visual Comparison of the extracted PRNUs with CNNs trained with different strategies. (a) The image, (b) DnCNN, PCE=206, (c) DnCNN-SPN-MSE, PCE=257, (d) DnCNN-SPN-Rho, PCE=256.

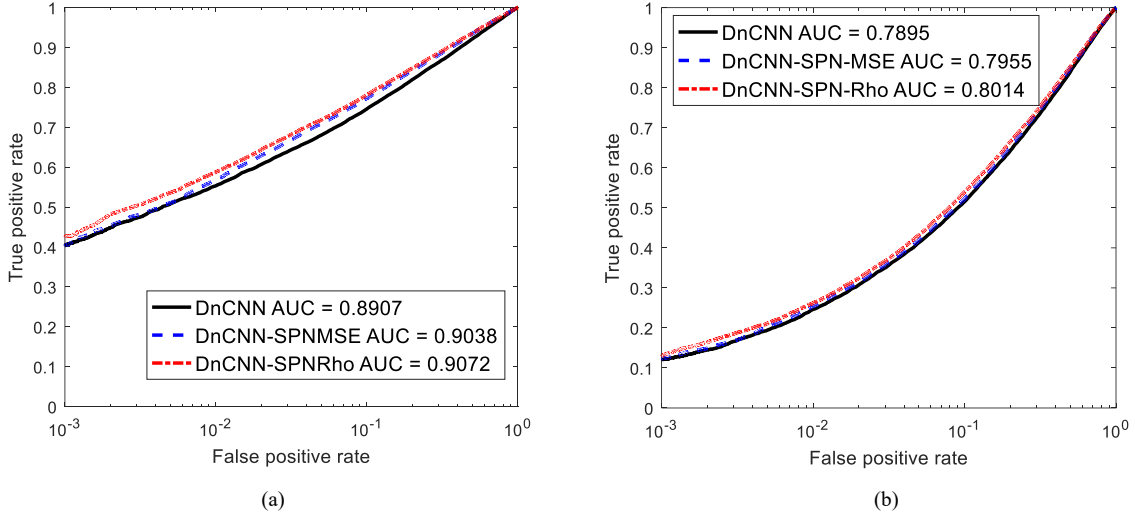

**Fig. 5.** ROC comparison of different training strategies. (a) 128 × 128 images, (b) 64 × 64 images.

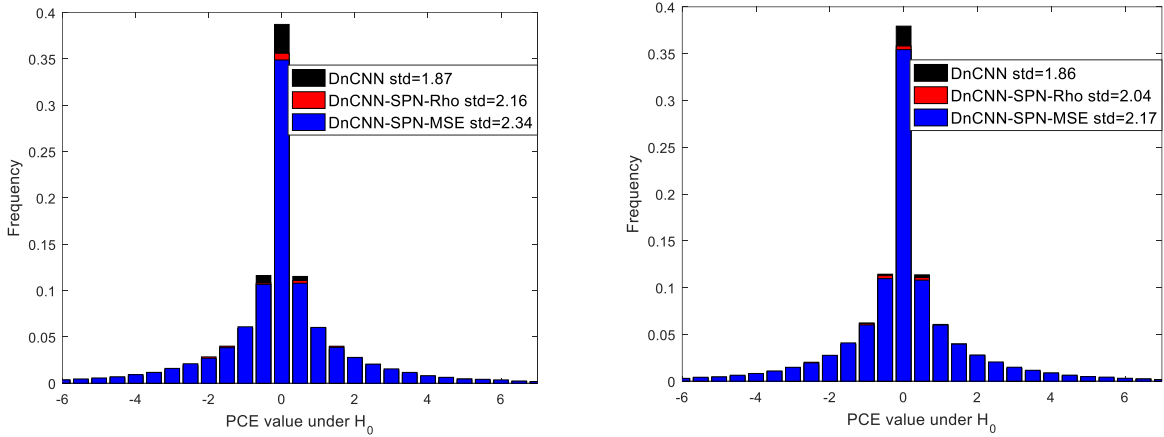

**Fig. 6.** Comparison of the distribution of PCE values under $H_0$. (a) 128 × 128 images, (b) 64 × 64 images.

PRNU pairs instead of image-synthetic noise pairs. Fig. 4 provides an intuitive comparison between the PRNU estimated with the CNNs trained with different strategies. This example shows that the PRNU estimated with DnCNN-SPN-MSE/DnCNN-SPN-Rho is rather free of image content, unlike the estimate obtained with DnCNN.

We also compare these three training strategies with ROC curves in Fig. 5. Again, training with image-PRNU pairs shows an advantage in terms of the AUC value. However, this advantage becomes marginal compared to the results in Table 3, especially for the case of DnCNN-SPN-MSE. Under careful examination, we find this is because the dynamic range of PCE values from DnCNN-SPN-MSE (or DnCNN-SPN-Rho) for negative samples is larger than that of DnCNN. Fig. 6 compares the distribution of the negative samples for different training strategies. PCEs for the negative samples with DnCNN-SPN-MSE and DnCNN-SPN-Rho are more "widely" distributed than those of DnCNN, thus making them easier to overlap with PCE distribution for the positive samples. For binary classification, not only the inter-class distance but also the intra-class distance determines the classification performance. According to our experiments, training CNN denoisers with image-PRNU pairs can significantly boost the inter-class distance between PCE distributions for positive and negative samples in SCI. However, it also runs the risk of enlarging the intra-class distance for negative samples, which is also of concern. Among the two networks trained with image-PRNU pairs, DnCNN-SPN-Rho outperforms DnCNN-SPN-MSE in all tests, which verifies that the proposed $\rho$-based loss function is a better alternative to MSE loss function in training PRNU extractors. This is a direct consequence of the detection function PCE defined as an increasing function of $\rho$. Notice that the network trained with the $\rho$-based loss function can implicitly use different strengths in extracting PRNU signals for different images. It is widely known that the strength of PRNU signals in images is affected by many factors, such as ISO settings and scene content, and varies from image to image. Hence, the ability to vary the strength of PRNU estimate implied in the $\rho$-based loss function may contribute to the better PRNU extraction performance.

Table 4 Time comparison of extracting PRNU from an image of size 1024×1024. CPU/GPU

| Method | Running time ($s$) |
|---|---|
| *DWT* [7] | 0.47/- |
| *DTCWT* [14] | 1.12/- |
| *BM3D* [13] | 6.85/- |
| *DnCNN* [22] | 7.62/1.02 |
| *FFDNet* [23] | 2.05/0.24 |
| *ADNet* [24] | 10.52/1.48 |
| *DANet* [26] | 3.15/0.46 |

4.4 Computation efficiency

Finally, we compare the time complexity of the PRNU extraction methods for images of size 1024×1024. Average running times are listed in Table 4. For the non-data-driven methods, the simulations are performed in MATLAB 2016a on a laptop with a 2.6 GHz Intel Core i7 CPU and 8 GB RAM. For the CNN denoisers, we report their efficiency both on the CPU platform and GPU platform with the Pytorch package [38]. The GPU we used is an Nvidia Quadro P600 with 4GB of graphic memory.

Since the simulations are performed on different platforms, we cannot directly compare non-data-driven methods and data-driven methods. According to the results in Table 4, CNN denoisers appear to be slower than non-data-driven methods except for BM3D when running on the CPU. However, when running the CNN denoisers on the GPU, they are generally faster than the non-data-driven methods.

5 Conclusions

This paper presents a comparison study for adaptions of the state-of-the-art CNN denoisers for PRNU estimation. We highlight the applicability of image denoising for PRNU extraction. Tailored strategies for training the CNN-based PRNU extractors are also introduced and evaluated in this work. We summarize our findings as follows:

1) The state-of-the-art CNN denoisers with hyper-parameters carefully chosen for PRNU estimation achieve comparable performance with the non-data-driven PRNU extractors, such as DTCWT and BM3D. The test results suggest that both the best data-driven methods and the best non-data-driven methods approach an upper limit for PRNU estimation.

2) The new strategy of training with image-PRNU pairs improves the PRNU estimation performance compared to feeding image-synthetic Gaussian noise pairs into CNNs. However, this new strategy may also increase false alarms in SCI by enlarging the range of metrics for negative samples.

3) The proposed $\rho$-based loss function can further improve the CNN-based PRNU extractor compared to that with the MSE loss function.

We also introduced a PRNU quantization scheme that can significantly reduce the storage requirements for camera fingerprints without impacting the correlation-based detector performance. The quantized reference PRNUs of DID, as well as all the trained PRNU extractors, are available on our GitHub for the convenience of future comparative studies.[1]

CRediT authorship contribution statement

**Hui Zeng:** Conceptualization, Methodology, Writing - original draft. **Morteza Darvish Morshedi Hosseini:** Conceptualization, Methodology, Writing - review & editing. **Kang Deng:** Validation. **Anjie Peng:** Data curation. **Miroslav Goljan:** Supervision, Writing - review & editing.

---

[1] The codes and quantized camera fingerprints are available at https://github.com/zengh5/PRNU_CNN.

## Acknowledgements

Special thanks belong to Dr. C. Tian for the discussion about CNN denoisers. This work is supported by NSFC (No. 61702429), Doctoral Research Fund of SWUST (No.18zx7163), China Scholarship Council (No. 201908515095), and DARPA and Air Force Research Laboratory under the research grant number FA8750-16-2-0173 and FA8750-16-2-0192.